\documentclass[sigconf]{acmart}
\AtBeginDocument{%
  \providecommand\BibTeX{{%
    \normalfont B\kern-0.5em{\scshape i\kern-0.25em b}\kern-0.8em\TeX}}}

\setcopyright{iw3c2w3}
\copyrightyear{2020}
\acmYear{2020}
\acmDOI{10.1145/1122445.XXXXXXX}

\acmConference[WWW '20]{Proceedings of The Web Conference 2020}{April 20--24, 2020}{Taipei, Taiwan}
\acmBooktitle{Proceedings of The Web Conference 2020 (WWW '20), April 20--24, 2020, Taipei, Taiwan} 
\acmPrice{}
\acmISBN{978-1-4503-7023-3/20/04}

\usepackage{subcaption}
\usepackage{xcolor}

\usepackage{titlesec}
\titleformat{\subsubsection}[runin]
{\normalfont\normalsize\bfseries}{\thesubsubsection}{1em}{}

\begin{document}

\title{Games for Fairness and Interpretability}

\author{Eric Chu}
\authornote{Both authors contributed equally.}
\email{echu@mit.edu}
\affiliation{%
  \institution{Massachusetts Institute of Technology}
}

\author{Nabeel Gillani}
\email{ngillani@mit.edu}
\authornotemark[1]
\affiliation{%
  \institution{Massachusetts Institute of Technology}
}

\author{Sneha Priscilla Makini}
\email{snehapm@mit.edu}
\affiliation{%
  \institution{Massachusetts Institute of Technology}
}

\renewcommand{\shortauthors}{Trovato and Tobin, et al.}

\begin{abstract}
As Machine Learning (ML) systems becomes more ubiquitous, ensuring the fair and equitable application of their underlying algorithms is of paramount importance. We argue that one way to achieve this is to proactively cultivate public pressure for ML developers to design and develop fairer algorithms --- and that one way to cultivate public pressure while simultaneously serving the interests and objectives of algorithm developers is through gameplay. We propose a new class of games --- ``games for fairness and interpretability'' --- as one example of an incentive-aligned approach for producing fairer and more equitable algorithms.  Games for fairness and interpretability are carefully-designed games with mass appeal. They are inherently engaging, provide insights into how machine learning models work, and ultimately produce data that helps researchers and developers improve their algorithms.  We highlight several possible examples of games, their implications for fairness and interpretability, how their proliferation could creative positive public pressure by narrowing the gap between algorithm developers and the general public, and why the machine learning community could benefit from them.
\end{abstract}

\ccsdesc[500]{Human-centered computing}

\keywords{machine learning, interpretability, fairness, games, crowdsourcing}

\maketitle

\section{Introduction}

As ML increasingly permeates virtually all aspects of life --- and unequally serves, or fails to serve, certain subsegments of the population \cite{caliskan2017semantics,bolukbasi2016man,genderShades} --- there is a need for a deeper exploration of how ML algorithms can be made fairer and more interpretable. To achieve this, we believe effective public pressure will be one lever to better models. There are several examples from history of how public pressure has spurred changes to technology policies. The creation of dynamite; America's use of the atomic bomb during the second world war; and the eugenics movement from the early 20th century are all examples of ethically dubious endeavors that were at least somewhat abated by a critical public response\footnote{https://www.bostonglobe.com/ideas/2018/03/22/computer-science-faces-ethics-crisis-the-cambridge-analytica-scandal-proves/IzaXxl2BsYBtwM4nxezgcP/story.html}.

However, recent stories about Facebook and Cambridge Analytica, driverless cars going rogue\footnote{https://www.nytimes.com/2018/03/23/technology/uber-self-driving-cars-arizona.html}, and even machine-powered labor displacement \cite{autorJobs} have hinted at the dangers of simply letting history unfold. In all of these instances, there were certainly changes to the underlying technological methods --- but it is hard to deny the importance of collective public pressure in catalyzing dialogue to envision a new set of policies and practices surrounding these powertools. It is unlikely that methodological changes alone would have been sufficient. Public pressure is often reactive and arises in the wake of crises. To counter this, we ask: how can public pressure operate proactively in order to ensure ML can effectively ground itself in --- and respond to --- calls for fairness and interpretability?

To that end, some authors have recently sparked public conversation around the ethical pitfalls of machine learning \cite{weaponsMath, automatingInequality,algosOppression}. Furthermore, initiatives like Turingbox \cite{turingBox} and OpenML \cite{vanschoren2014openml} are actively seeking to create platforms and marketplaces where members of the scientific community and general public can audit ML algorithms to promote more fairness, transparency, and accountability.  These efforts are important first steps towards generating proactive public pressure. However, they fail to directly align incentives between those who design and deploy algorithms and those who are affected by them.  Why should an algorithm developer care about how a niche group of individuals rates the fairness or interpretability of his or her algorithms?  Why should members of the general public spend their time trying to understand, let alone evaluate, these algorithms?  It is unclear how sustainable current efforts to generate proactive public pressure will be without incentive alignment.

To align incentives between ML developers and the general public in a quest for more interpretable --- and as a result, in due course, fairer --- ML, we propose ``games for fairness and interpretability'': networked games that as a byproduct of the game's objectives, engage the general public in auditing algorithms while simultaneously generating valuable training sets for ML developers.

\section{ML Powered Games}

Inspired by Luis von Ahn's Games with a Purpose (GWAP) framework \cite{von2008human,von2008designing}, we propose using ML-powered games to enhance model interpretability --- which we view as an important step towards developing fairer ML.

\subsection{Games with a Purpose}
Described as ``human computation", the GWAP framework was designed for problems solvable by humans but beyond the capabilities of machines. Instead of relying on financial incentives or altruism, GWAPs simply rely on people's desire for fun and entertainment. A successful GWAP can produce not only novel and creative solutions to difficult problems, but also provide large amounts of labeled data for training machine learning models. Since its inception, GWAPs have attracted hundreds of thousands of players in order to tackle problems ranging from protein folding \cite{khatib2011crystal} and RNA folding \cite{lee2014rna} to examining the human perception of correlation in scatter plots\footnote{http://guessthecorrelation.com/}. The framework has also since been extended to machine learning, such as using active learning to select examples during gameplay \cite{barrington2012game}.

\begin{figure}[h]
  \centering
  \includegraphics[width=\linewidth]{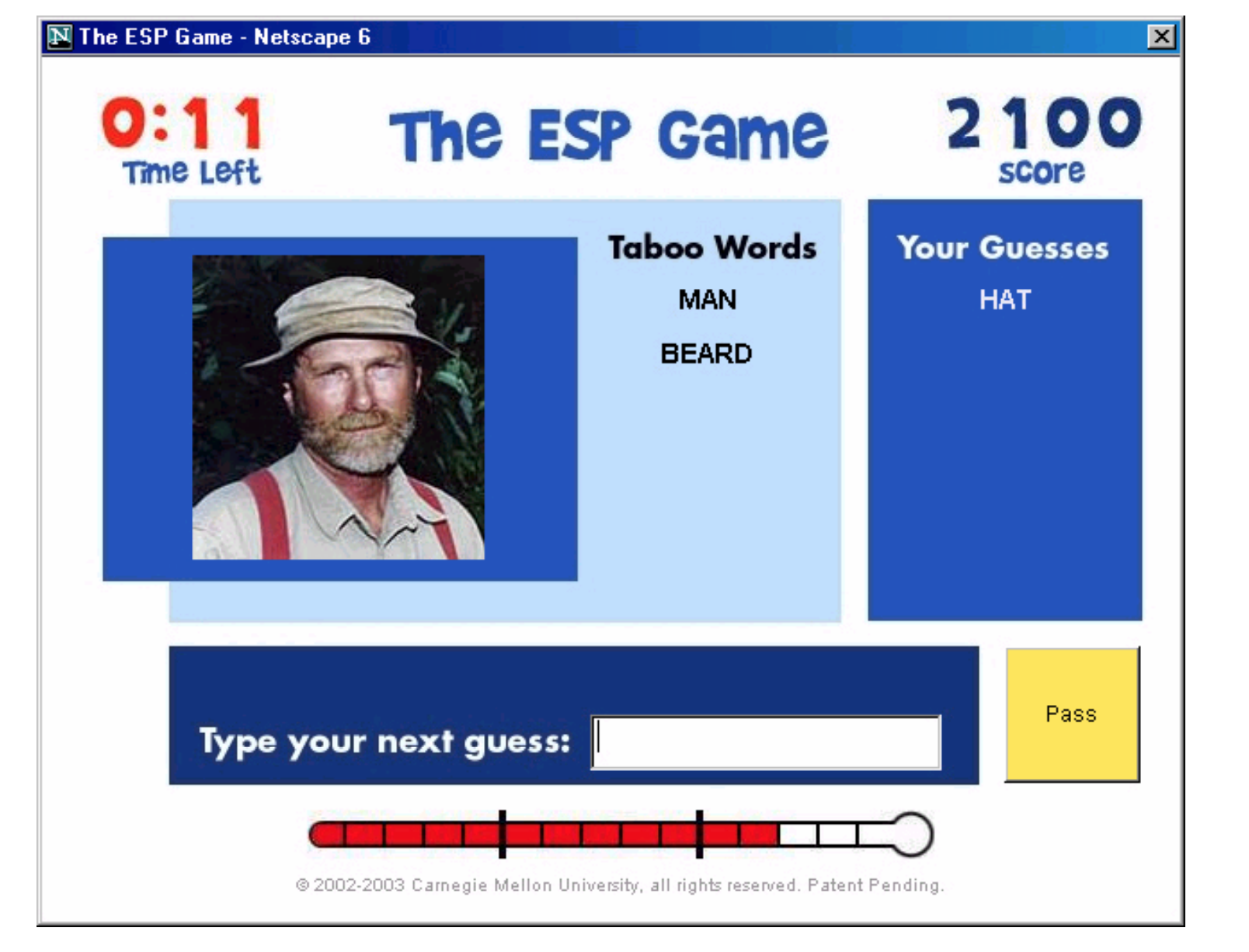}
  \caption{An example of a Game With a Purpose (GWAP): the original ESP game.}
  \label{fig:esp}
\end{figure}

The GWAP framework includes several different templates of games \cite{von2008designing}. \textit{Output-agreement} games has two players attempt to produce the same output when shown the same input. In the ESP game, shown in Figure \ref{fig:esp}, the players are shown an image and asked to guess what words the other player would use to describe the image. A variation of the game includes taboo words for each image, thus requiring users to guess more uncommon words, in turn producing more interesting labeled data \cite{von2004labeling}. In \textit{input-agreement} games, two players are each provided an input which may or may not be different; the players are asked to output descriptions of the inputs and then finally guess whether they were shown the same input. For instance, players in the Tagatune game are given song clips and asked to output tags, before finally guessing whether they had the same clip \cite{law2009input}.

%

\subsection{Designing Games for Fairness and Interpretability}

While reputation-based incentives can create social pressure and motivate ML developers, we believe a well-designed game aligns incentives between ML developers and the consumers of ML (i.e. the general public). Due to the importance of labeled data for deep neural networks, we believe ML researchers will have strong incentives to upload their models if the games that leverage them can produce valuable training data or adversarial examples.

On the consumer side, GWAPs have shown that such games can reach large audiences. Furthermore, a larger audience is often a broader audience, thus allowing more diverse probing of the model. We believe that there is an appetite for ML games, due both to increasing media attention on ML and the growing capabilities of new models. Recent examples of games that engage a general audience in exploring ML include the text auto-complete ``Talk to Transformer''\footnote{https://talktotransformer.com/}, a Pictionary-like game Quick, Draw!\footnote{https://quickdraw.withgoogle.com/}, word embedding-powered word association games\footnote{http://robotmindmeld.com/}, and an endless text-adventure game built using a generative text model\footnote{https://www.aidungeon.io/}.



We define ``games for fairness and interpretability'' as ML-powered games in which the output and / or interaction with human players is produced by a machine learning model. These games can also be networked to enable human-human interaction and competition. Games should be fun and engaging, provide insight into how the underlying machine learning models work, and produce data that helps models improve --- in particular, so that the models are better-equipped to more equitably serve a diverse range of individuals and scenarios. 

One might imagine a platform for such games, where once a game has been designed and open-sourced, its backend model could be swapped for any model with similar inputs and outputs. The platform could also serve as a public forum for widespread participation in, and discussion about, the evaluation of new ML models. This unique forum --- one where both ML developers and members of the public are present --- could serve as an important vehicle for a) enhancing broader familiarity with and awareness of ML and its applications, and perhaps eventually, b) creating proactive public pressure that motivates algorithm developers to build more interpretable and fairer ML.



\begin{figure*}[!h]
    \centering
    \begin{subfigure}{0.5\textwidth}
        \centering
        \includegraphics[width=0.85\textwidth]{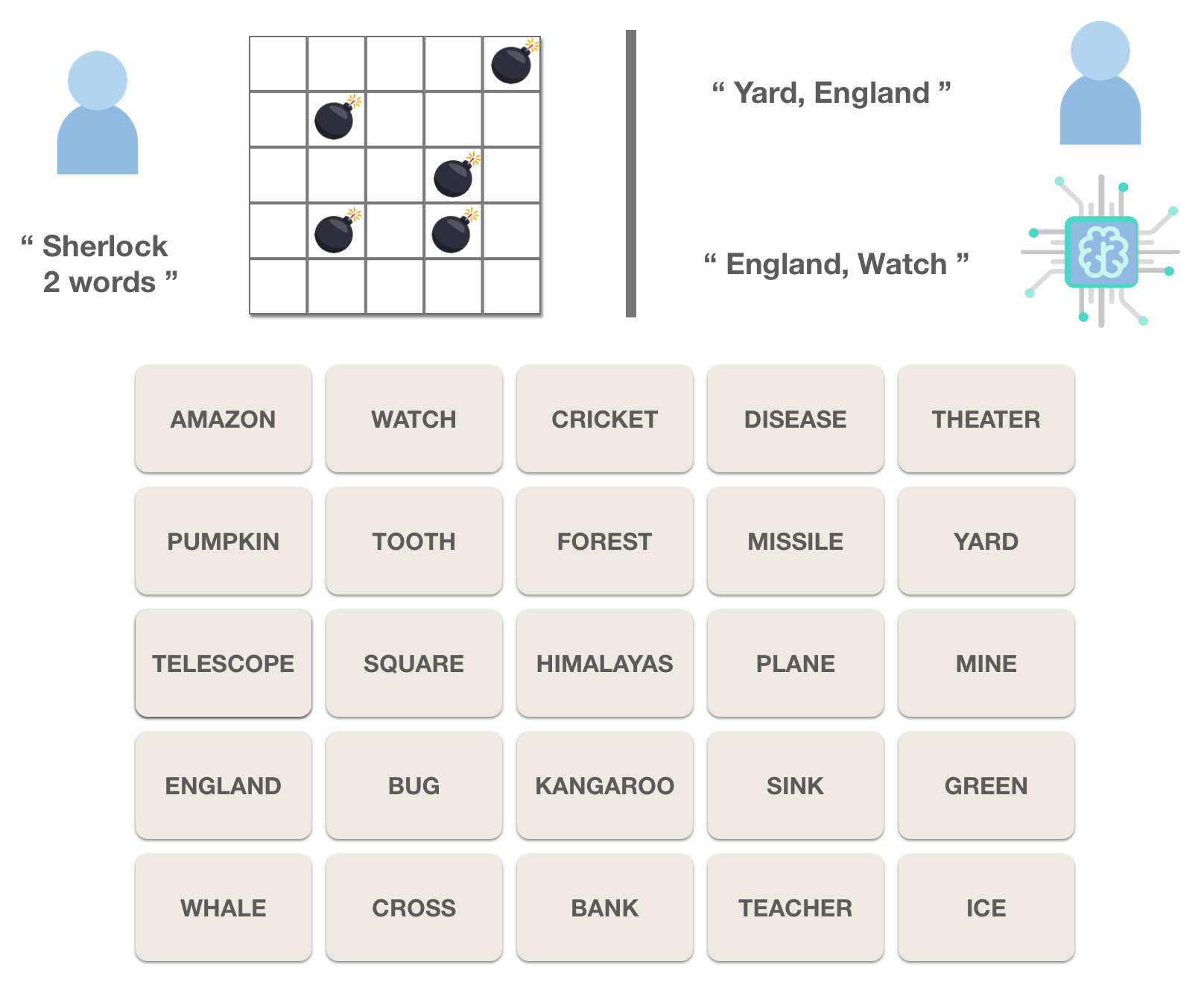}
        \caption{Example of a \textit{Humans vs. AI} game. Player 1 (the ``spymaster'') provides an input, while Player 2 competes against an AI to produce the correct answer. Here, since the human and the AI both guessed ``England'', only ``Yard'' would count as a correct answer.}
        \label{fig:codenames}
    \end{subfigure}%
    \begin{subfigure}{0.5\textwidth}
        \centering
        \includegraphics[width=0.8\textwidth]{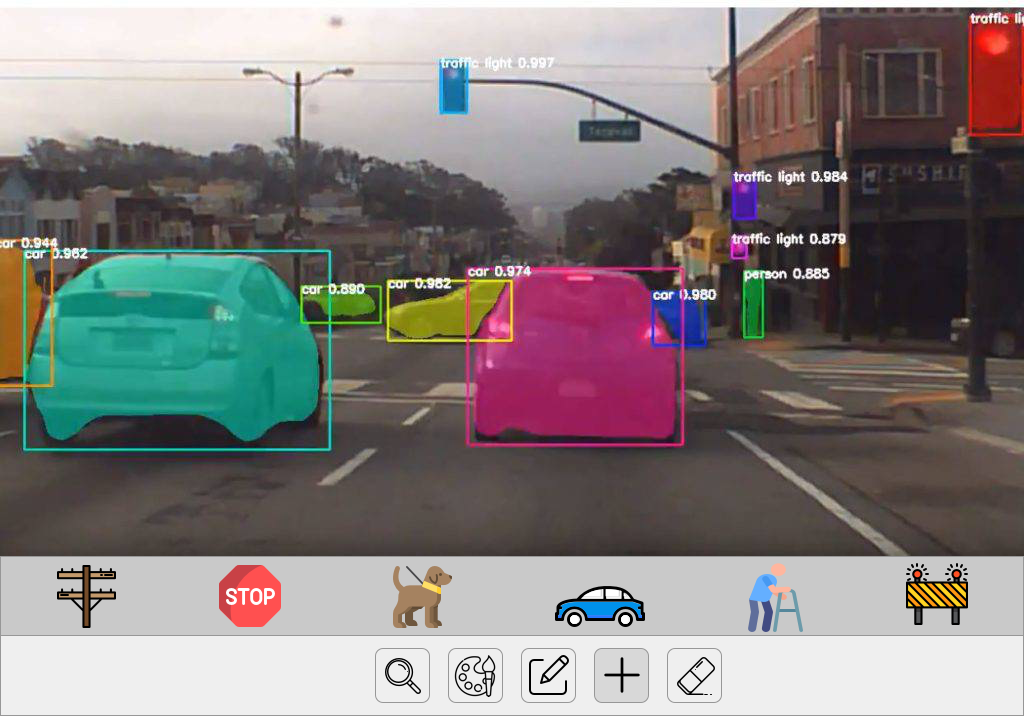}
        \caption{Example of a \textit{Break the Bot} game. Player 1 and Player 2 compete against each other in producing adversarial attacks that will reduce the accuracy of the model's predictions. Players should be incentivized to make small edits that nevertheless produce large decreases in accuracy. In this example, players are provided tools to change the lighting and color, or add and remove common objects.}
        \label{fig:viz}
    \end{subfigure}
    \caption{Examples of possible Games for Fairness and Interpretability. Both types of games are designed to surface model biases and deficiencies, while also producing more robust and diverse training data.}
\end{figure*}

\subsection{Proposed Categories of Games}
In the spirit of GWAPs, we describe possible categories of games in the following sections.

\subsubsection{Humans vs. AI}

\hfill \break
\indent\textbf{Setup.} Player 1 provides an input, and Player 2 competes against an AI to guess the correct answer.

\textbf{Example game 1 --- Guess Who?} Player 1 describes themselves, their interests, job, and other attributes through freeform short text. Player 2 and the AI attempt to guess the age, sex, and location of Player 1.

\textbf{Example game 2 -- Codenames}. Inspired by the popular Codenames board game \cite{wiki:codenames} , the players are presented with a 5x5 grid of words. Player 1 is a ``spymaster'' who is also allowed to see the placement of bombs on the grid. The spymaster's role is to give a one word clue, plus the number of words that matches the clue. Player 2's goal is to guess the correct words; however, if he or she guesses a bomb, the game is over. The game is won if all the non-bomb words are guessed correctly. The goal is to finish the game in fewer rounds; saying a larger number allows the team to win more quickly, but it is also more difficult to come up with clues. 

In our ML-powered variant, the AI also attempts to guess the words; if the AI's guesses matches Player 2's guesses, those guesses are invalid. Figure \ref{fig:codenames} shows an example round. 


\textbf{Data produced and insight into interpretability.} Player 1 will have to produce inputs that are recognizable by another human but undetectable or incorrectly classified by the AI. This requires a player to intuit the space of inputs that a model understands and in which cases it might fail. For instance, Player 1 may find that cultural references are harder for a ML model. Natural language processing models that can incorporate common sense reasoning and knowledge also remains an open area of research. The successful inputs and clues can be used as more robust training data. 

In addition, baseline models for the AI could be based on word embeddings, which have been shown to reflect implicit human biases around gender, race, occupation, etc. \cite{caliskan2017semantics}. These biases may be surfaced if the AI incorrectly relies on them to make predictions.

\subsubsection{Break the Bot}

\hfill\break
\indent\textbf{Setup.} Each player is shown an input and the model's output (e.g. a prediction). Each player is asked to make a small modification to the input. Whoever can cause the largest change in the model output, while using the smallest modification, receives more points.

\textbf{Example game 1 --- Vandalize it!} The brittleness of deep neural networks has been illustrated in several computer vision systems. For example, graffiti on signs can significantly lower object recognition accuracy \cite{eykholt2018robust}, while Rosenfeld et al. showed that adding an object to a scene could drastically change the ability to recognize all other objects \cite{rosenfeld2018elephant}. These deficiencies can have catastrophic effects on real-world systems.

In this self-driving car inspired game, players are shown street images overlaid with bounding boxes of detected objects. For example, a stop sign may be detected by the model with probability 0.85. The players' goal is to change that prediction by making small edits to the sign and its surroundings. The game will give players tools to alter the angle, lighting, hue of the image, as well as add and subtract other objects and artificats. (The game will have to measure the `size' of modifications in order to assign scores). Figure \ref{fig:viz} shows an example of how the game might look.

\textbf{Example game 2 --- Beat the Banker.} ML has begun to be used in higher-stakes situations, ranging from recidivism prediction to loan default rate prediction. Unfortunately, these systems have also been shown to be susceptible to demographic features and unfairness \cite{hardt2016equality}. In this game, the players are bankers. The input is a hypothetical set of demographic features of an individual, and the output is the predicted probability of that individual's loan repayment. Faced with a loan rejection, the goal is to find seemingly innocuous changes that can make the loan approved.



\textbf{Data produced and insight into interpretability.} 
These games provide adversarial examples and sensitivity analysis on model inputs. This is important as the field of adversarial examples is becoming increasingly important \cite{goodfellow2014explaining}, especially as ML models become deployed in the real world \cite{kurakin2016adversarial}, and obtaining those examples can often be difficult \cite{zhao2017generating}. ML researchers can also gain a greater understanding of how inputs may be modified in semantically meaningful ways, as well as if the observed model behavior is desirable (e.g. fair).

\section{Games and Current Research Directions in Machine Learning}
The previous section illustrates how thoughtfully-designed games might help align incentives between ML developers and the general public, cultivating public pressure and awareness --- along with the new, more representative datasets --- to promote fairer, more inclusive ML systems.  We believe the time to develop games for fairness and interpretability is now, largely because they align with several current directions in ML research. We highlight some of these directions below and explore how members of these respective research communities may benefit from games for fairness and interpretability.

\subsection{Fairness}
As ML models become more pervasive, there has been an increasing call for models that can prevent discrimination along sensitive attributes such as race and gender. Part of the problem is detecting that biases in models even exist in the first place. To that end, recent research has shown how word embeddings encode biases as measured by standard tests such as the Implicit Association Test \cite{caliskan2017semantics}, with relationships between word embeddings reflecting negative stereotypes about gender \cite{bolukbasi2016man}. Other work highlights deficiencies in datasets used for facial recognition, resulting in models that fail more frequently for women and people with darker skin tones \cite{genderShades}.  

How can models handle these sensitive attributes? A naive approach of removing sensitive attributes may not prevent discrimination if the sensitive attributes are correlated with other attributes left in the dataset. Enforcing demographic parity, in which the outcome is uncorrelated with the sensitive attribute, is also problematic because it does not guarantee fairness, and the sensitive attribute may actually be important for prediction, making removal of all correlation unrealistic. Thus far, various approaches to formalize and operationalize fairness include using the 80\% rule of ``disparate impact'' outlined by the US Equal Employment Opportunity Commission as a definition of discrimination \cite{feldman2015certifying}, treating similar individuals similarly by enforcing a Lipschitz condition on similar individuals and the classifier predictions for those individuals \cite{dwork2012fairness}, preprocessing the dataset through methods such as weighting and sampling \cite{kamiran2012data}, and allowing use of the sensitive attribute but aiming for ``equality of \textit{opportunity}'' through the notion of equalized odds \cite{hardt2016equality}. Certain frameworks also provide the ability for people to select the tradeoff between model performance and fairness. Other work has centered on learning transferable\textit{ fair }representations that can be reused across tasks \cite{zemel2013learning}. 

We believe ML fairness researchers would find value in the datasets produced by games for fairness and interpretability.  For example, machine predictions from ``Human vs. AI'' games would provide clear insights into which kinds of biases certain algorithms harbor; ``Break the Bot'' games might shed light on how robust or brittle algorithms are to changes in the datasets they operate on. 

\subsection{Interpretability}
While deep neural networks have found great success as powerful function approximators, they have also developed a reputation as black boxes. Interpretability may be a case of ``you know it when you see it'', but recent work has attempted to make the problem more tractable by defining interpretability, explaining why it is important, and explaining when it is necessary \cite{doshi2017towards, lipton2016mythos}. 


There has also been a wide range of methods focused on \textit{introspection and visualization}, including (but not limited to) ``inverting'' intermediate representations to generate images \cite{mahendran2015understanding, mordvintsev2015inceptionism},  producing input feature attributions and saliency maps \cite{ribeiro2016should,sundararajan2017axiomatic,smilkov2017smoothgrad,petsiuk2018rise,bau2017network} vs. producing counterfactual explanations \cite{wachter2017counterfactual} vs. pointing to protoypical examples \cite{chen2019looks}, local per-example explanations \cite{ribeiro2016should} vs. global explanations based on feature representations across the entire dataset \cite{bau2017network}, clear-box approaches with access to model gradients \cite{sundararajan2017axiomatic,smilkov2017smoothgrad} vs. black box approaches \cite{ribeiro2016should, petsiuk2018rise}. These methods often highlight what parts of the input (e.g. a segment of the image, or a span of the text), were most important to the model's decision.

While there is also work worth mentioning on (1) generating human readable explanations in natural langauge \cite{lei2016rationalizing, hendricks2016generating}, (2) distilling neural networks into more interpretable models such as decision trees \cite{frosst2017distilling}, and (3) disentangling factors of variation for generative models \cite{higgins2016beta,chen2016infogan, kulkarni2015deep}, a significant portion of the field has focused on the aforementioned introspection and visualization methods. At the core, many of the methods attempt to relate input or internal representations to the model outputs.  However, there are questions around the reliability and intuitiveness of these explanations (\cite{jain2019attention, kindermans2019reliability}). The games' data can be analyzed through these methods, perhaps providing insight into how well current explanations match human intuitions. The ``Break the Bot`` games would also produce valuable counterfactual data; analyzing changes in the outputs of their underlying models as a function of changes to inputs could provide a deeper understanding of how, exactly, these models are conducting their computations.

\section{Conclusion}
As ML-powered technologies continue to proliferate, the threat of biased and opaque decision-making looms large. We believe public pressure is a powerful mechanism for inspiring changes in how algorithms are developed. Games for fairness and interpretability provide one means for engaging the public in probes of ML systems while simultaneously producing hard-to-source data that serves the interests of ML developers. We believe games are unique in their ability to engage different audiences and are thus a promising avenue in which to pursue complicated, multi-stakeholder challenges like building fairer ML systems.


Looking ahead, there are several open questions: who should be responsible for designing and developing games for fairness and interpretability?  How will the games be deployed and marketed so as to recruit a diverse range of players?  What new risks or threats might these games introduce?  These are important questions that will require continuous exploration and reflection.  We hope this paper serves as an initial stepping stone and inspires individuals both within and beyond the ML community to consider the potential power of games.

\bibliographystyle{ACM-Reference-Format}
\bibliography{main}

\end{document}